\documentclass{aa}

\usepackage{graphicx}
\usepackage{amsmath}
\usepackage{mathtools}
\graphicspath{{figures/}}
\usepackage{txfonts}
\begin{document}

   \title{Radio timing in a millisecond pulsar-extreme/intermediate mass ratio binary system}

   \author{T. Kimpson
          \and 
          K. Wu 
          \and
          S. Zane
          }
   \institute{Mullard Space Science Laboratory, University College London. Holmbury St. Mary, Dorking, Surrey, RH5 6NT, UK \\
   	\email{t.kimpson.16@ucl.ac.uk}
             }

   \date{Received XXX; accepted YYY}

 
\abstract{Radio timing observations of a millisecond pulsar in orbit around the Galactic centre black hole (BH) or a BH at the centre of globular clusters could answer foundational questions in astrophysics and fundamental physics. Pulsar radio astronomy typically employs the post-Keplerian approximation to determine the system parameters. However, in the strong gravitational field around the central BH, higher order relativistic effects may become important. We compare the pulsar timing delays given by the post-Keplerian approximation with those given by a relativistic timing model. We find significant discrepancies between the solutions derived for the Einstein delay and the propagation delay (i.e. Roemer and Sharpiro delay) compared to the fully relativistic solutions. Correcting for these higher order relativistic effects is essential in order to construct  accurate radio timing models for pulsar systems at the Galactic centre and the centre of globular clusters and informing issues related to their detection.}


   \maketitle
%

\section{Introduction}
The gyroscopic nature of radio pulsars (PSRs) allows them to be used as exceptionally stable and regular clocks. By measuring the times of arrival (ToAs) of the pulsar pulses, an observer can determine information on the rotational and astrometric properties of the PSR, in addition to information on any binary companion \citep[e.g. ][]{Liu2011,Desvignes2016,Lazarus2016,Liu2018,Parthasarathy2019}. In the Galactic centre, there is expected to be a large population (up to $\sim10^4$ within the central parsec) of millisecond pulsars (MSPs)  \citep{Wharton2012, Rajwade2017}, whilst the number of pulsars in globular clusters is observed to be greater by a factor of $10^3$ per unit mass compared to the Galactic disk \citep{Freire2013}. In addition to MSPs, the Galactic centre is known to host a supermassive black hole (BH) with a mass of $M \sim 4 \times 10^6 M_{\odot}$\citep{Gillessen2009, Boehle2016}, whilst consilient strands of evidence from stellar kinematics \citep[e.g. ][]{Gebhardt2002, VDM2010, Feldmeier2013}, radiative accretion signatures \citep[e.g. ][]{Ulvestad2007} and pulsar dynamics \citep{K2017NAT} all suggest that globular clusters could host intermediate mass black holes (IMBHs), although definitive `smoking gun' evidence is still lacking, and some evidence is disputed \citep{Miller2012,Baumgardt2017}. Although no MSPs  have currently been detected in the Galactic centre or at the centre of globular clusters, the increased sensitivity of next generation of radio facilities - such as the Square Kilometer Array \citep[SKA,][]{Keane2015}, the Five-hundred-meter Aperture Spherical Telescope \citep[FAST,][]{Nan2011}, or the NASA Deep Space Network \citep[DSN, ][]{Majid2019}  - will make this population accessible. \newline 

 Radio timing the orbit of an MSP around a much more massive BH has been identified as an excellent apparatus with which to probe general relativity (GR) and astrophysics in a highly strong-field, non-linear regime \citep{Kramer2004, Liu2012, Psaltis2016}. These systems are known as extreme (mass ratio $q \sim 10^6$) or intermediate ($q \sim 10^3 - 10^4$) mass ratio binaries (E/IMRBs). Via radio timing of an MSP-E/IMRB, the essential parameters of the central BH - the mass, spin, and quadrupole moment ($M, S, Q$ respectively) -  can be determined to remarkable precision \citep[e.g.][]{Liu2012,Psaltis2016}. Once the BH parameters have been established, it is then possible to explore fundamental questions of GR such as the cosmic censorship conjecture \citep{Penrose1969} and no hair theorem \citep{Israel1967}. From the perspective of astrophysics, one could then establish whether the central dark nuclei of star clusters are indeed astrophysical BHs as described by the Kerr solution, or else something more exotic such as a boson star \citep{Kleihaus2012} or a BH described by some deviation from the Kerr solution  \citep[`bumpy' black holes][]{Yagi2016}. Moreover, if an MSP could be detected orbiting an IMBH, this would also naturally settle the debate on the existence of intermediate mass, astrophysical BHs \citep{Singh2014}. \newline

In order to determine the system parameters via radio timing, it is necessary to have a timing model which maps the time at which the pulse was received in the observers frame $t$ to the time at which the pulse was emitted in the `clock frame' $\tau$ (i.e. the frame comoving with the pulsar). For pulsars in binary systems, there are four key delays which must be accounted for in the timing model:
\begin{eqnarray}
t = \tau + \Delta_E + \Delta_R + \Delta_S 
\end{eqnarray}
where $\Delta_R$ is the Romer delay due to orbital motion of the pulsar, $\Delta_S$ is the Shapiro delay due to gravitational time dilation, and $\Delta_E$ is the Einstein delay, the relativistic time dilation \citep[see][for further details]{Lorimer2004}. The relationship between the timing delays ($\Delta_R ,\Delta_S , \Delta_E $) and the system parameters is based on the `Post-Keplerian' (PK) approximation \citep{Blandford1976, Damour1986} and is derived under the assumption that the gravitational field is weak. GR is subsequently treated perturbatively as an expansion about Newtonian gravity. Whilst this considerably simplifies the previously non-linear equations, such an approximation may not be appropriate for relativistic orbits close to massive BHs. The strength of the gravitational potential can be quantified via the parameter 
\begin{eqnarray}
\epsilon = \frac{GM}{rc^2} \ .
\end{eqnarray}
Tests of GR in the solar system which employ the PK framework are undertaken in an environment where typically $\epsilon \sim 10^{-8}$.  In binary pulsar systems $\epsilon \lesssim 10^{-6}$. Conversely, for a pulsar at periapsis on a $P = 0.1$ year eccentric ($e=0.9$) orbit around Sgr A$^*$, $\epsilon = 0.01$. Due to external perturbations on the PSR orbit \citep{Merritt2010}, the cleanest information on the BH parameters will be obtained from the PSR at periapsis, in precisely the region of strongest gravity. Photons propagating from the pulsar to an observer may pass sufficiently close to the BH so as to traverse a spacetime where $\epsilon$ approaches unity \citep[see e.g. discussion on the use of pulsars to probe quantum effects and the BH event horizon][]{Estes2017}. It therefore seems reasonable to question the validity of the PK weak-field approximation in a region where the gravitational potential is $\gtrsim 10^4$ times stronger than the environments in which it is typically applied. \newline 

Pulsar timing delays given by the PK approximation in strong-field regimes was partly addressed in \citet{Hackmann2018} who investigate the propagation delays (Roemer and Shapiro) from circular orbits in a Schwarzschild spacetime. The influence of the central black hole spin on the measurement of pulsar orbital parameters was investigated in \citet{Zhang2017}, whilst \citet{Li2019} explored the impact of relativistic spin dynamics on the PSR emission. In this work, we further develop these previous studies by deploying a consistent, relativistic framework \citep{Kimpson2019a, Kimpson2019b} to accurately calculate the pulse ToA from a spinning pulsar orbiting a Kerr BH. We can then compare this `true' GR solution with the `approximate' PK solution and so determine the validity of the PK approach for BH parameter estimation at the Galactic centre and the centre of globular clusters. \newline 

This paper is organised as follows. In Section \ref{sec:RTM} we review the construction of a relativistic pulsar timing model via numerical methods. In Section \ref{sec:PK} we overview the PK parameters that are used for tests of GR and compare the PK solution with the GR solution. Discussion and concluding remarks are made in Section 4. We adopt the natural units, with $ c=G=\hbar = 1$, and a $(-,+,+,+)$ metric signature.  Unless otherwise stated, a c.g.s. Gaussian unit system is used in the expressions for the electromagnetic properties of matter. The gravitational radius of the black hole is $r_{\rm g} = M$ and the corresponding Schwarzschild radius is $r_{\rm s} = 2M$, where $M$ is the black-hole mass. A comma denotes a partial derivative (e.g.$\;\! x_{,r}$), and a semicolon denotes the covariant derivative (e.g.$\;\! x_{;r}$).

\section{Relativistic MSP Timing Model}
\label{sec:RTM}
We now briefly review the methods of \citet{Kimpson2019b} for simulating the MSP timing signal. This has two primary components: the spin-orbital dynamics of a spinning MSP around a spinning BH, and the propagation of the MSP radio beam to the observer. We refer the reader to \citet{Kimpson2019b} for details regarding the algorithmic combination of these two effects to generate a consistent ToAs. For this work we ignore chromatic plasma effects and work solely in vacuum.

\subsection{Kerr spacetime}
The spacetime around a rotating BH is described by the Kerr solution. In Boyer-Lindquist coordinates, the metric interval is given by,
\begin{eqnarray}
{\rm d}s^2 = -\left(1 - \frac{2Mr}{\Sigma}\right) {\rm d}t^2 
- \frac{4aMr \sin^2 \theta}{\Sigma}\ {\rm d}t \;\! {\rm d}\phi 
+ \frac{\Sigma}{\Delta}{\rm d}r^2 + \Sigma\ {\rm d} \theta^2 \nonumber \\ 
\hspace*{1.2cm} + \frac{\sin^2 \theta}{\Sigma} \left[(r^2+a^2)^2 - \Delta a^2 \sin^2 \theta \right] {\rm d}\phi^2 \ , 
\label{eq:kerr_metric} 
\end{eqnarray}
where $\Sigma = r^2 + a^2 \cos^2 \theta$, $\Delta = r^2 - 2Mr +a^2$ and $a$ is the black-hole spin parameter. Generally we set the BH mass $M=1$ such that the lengthscale is normalised to the gravitational lengthscale $r_g$. The Kerr spacetime is both stationary -which leads to the conservation of energy $E$ -and axisymmetric - associated with the conservation of the projection of the angular momentum, $L_z$. In additional to these two conserved quantities which are associated with the Killing vectors of the spacetime, the Carter constant $Q$ \citep{Carter1968} is also conserved, which is associated with a Killing tensor. The Carter constant does not have an exact physical interpretation, but is related to the inclination of an orbiting body, and this is further discussed in \citet{DeFelice1999, Rosquist2009}. 

\subsection{Spin-orbital dynamics}
\label{section:MPD}
Astrophysical pulsars are not in reality point particles but extended spinning objects. This spin couples to the background spacetime and as a consequence a pulsar will not follow a geodesic of the Kerr metric. Instead, in limit of extreme mass ratio the MSP spin orbital dynamics can be described by the background gravitational field and the dynamical spin interaction with this field. The equations of motion of the pulsar are then determined via the MPD formulation \citep{Mathisson1937,Papapetrou1951, Dixon1974}. Via this approach one constructs the `gravitational skeleton' of the body by performing a multipole expansion of the stress energy tensor, $T^{\mu \nu}$. Since the general equation of motion is defined via,
\begin{eqnarray}
{T^{\mu \nu}} _{;\nu} = 0 \ ,
\end{eqnarray}
then to dipole order the equations of motion are,
\begin{eqnarray}
\frac{D p^{\mu}}{d \tau} = - \frac{1}{2} {R^{\mu}}_{\nu \alpha \beta} u^{\nu} s^{\alpha \beta} \ ,
\label{Eq:mpd1}
\end{eqnarray}
\begin{eqnarray}
\frac{D s^{\mu \nu}}{d \tau} = p^{\mu} u^{\nu} -p^{\nu} u^{\mu} 
\label{Eq:mpd2}
\end{eqnarray}
\citep{Mathisson1937,Papapetrou1951,Dixon1974}, where $p^{\mu}$ is the 4-momentum (mass monopole), $s^{\mu \nu}$ the spin dipole,  $u^{\nu}$ is the MSP 4-velocity and ${R^{\mu}}_{\nu \alpha \beta}$ the Riemann curvature tensor. We use the proper time $\tau$ to parameterise the MSP worldline and $D/d\tau $ denotes a covariant derivative. To make the system of equations determinate one must specify the reference worldline with which to define the multipole expansion, equivalent to choosing an observer with which with respect to which the centre of mass is defined, since in GR the centroid is observer-dependent. We  take the centroid to be that measured in  zero 3-momentum frame:
\begin{eqnarray}
s^{\mu \nu} p_{\nu} = 0 
\label{Eq:ssc}
\end{eqnarray}
\citep{Tulczyjew1959,Dixon1964}, the so-called TD condition. For a discussion on alternative centroid choices see \citet{Costa2014, Babak2014}. \newline

Given that the Moller radius of the MSP if much less than the physical radius, and the MSP mass much less than the BH mass, the pole-dipole interaction is much stronger than the dipole-dipole interaction and the equations of motion reduce to
\begin{eqnarray}
\frac{D u^{\mu}}{d \tau} = - \frac{1}{2m} {R^{\mu}}_{\nu \alpha \beta} u^{\nu} s^{\alpha \beta} \ ,
\end{eqnarray}
\begin{eqnarray}
\frac{D s^{\mu \nu}}{d \tau} \approx 0 
\end{eqnarray} 
\citep{Chicone2005,Mashhoon2006}. The complete set of ODEs is then
\begin{eqnarray}
\frac{dp^{\alpha}}{d\tau} = - {\Gamma^{\alpha}}_{\mu\nu} p^{\mu}u^{\nu} +  \left( \frac{1}{2m} {R^{\alpha}}_{\beta \rho \sigma} \epsilon^{\rho \sigma}_{\quad \mu \nu} s^{\mu} p^{\nu} u^{\beta}\right) \ ,
\label{eq:MPD1}
\end{eqnarray}

\begin{eqnarray}
\frac{ds^{\alpha}}{d \tau} = - {\Gamma^{\alpha}}_{\mu \nu} s^{\mu}u^{\nu} +  \left(\frac{1}{2m^3}R_{\gamma \beta \rho \sigma} \epsilon^{\rho \sigma}_{\quad \mu \nu} s^{\mu} p^{\nu} s^{\gamma} u^{\beta}\right)p^{\alpha} \ ,
\end{eqnarray}

\begin{eqnarray}
\frac{dx^{\alpha}}{d\tau} = -\frac{p^{\delta}u_{\delta}}{m^2} \left[ p^{\alpha} + \frac{1}{2} \frac{ (s^{\alpha \beta} R_{\beta \gamma \mu \nu} p^{\gamma} s^{\mu \nu})}{m^2 + (R_{\mu \nu \rho \sigma} s^{\mu \nu} s^{\beta \sigma}/4)}\right] 
\label{eq:MPD2}
\end{eqnarray}
 \citep{Singh2005,Mashhoon2006}, where $s^{\mu}$ is the spin 4-vector contraction of the spin tensor,
 \begin{eqnarray}
 s_{\mu} = -\frac{1}{2m} \epsilon_{\alpha \beta \mu \nu} p^{\nu} s^{\alpha \beta}
 \end{eqnarray}
Solving Eqs. \ref{eq:MPD1} - \ref{eq:MPD2}  numerically via a 5th order Runge-Kutta-Fehlberg algorithm \citep{Press1977} then completely specifies the spin-orbital dynamics of the pulsar, accounting for the background spacetime and the dynamical interaction of the MSP spin. 

\subsubsection{Orbital specification}
\begin{figure}
	\includegraphics[width=\columnwidth]{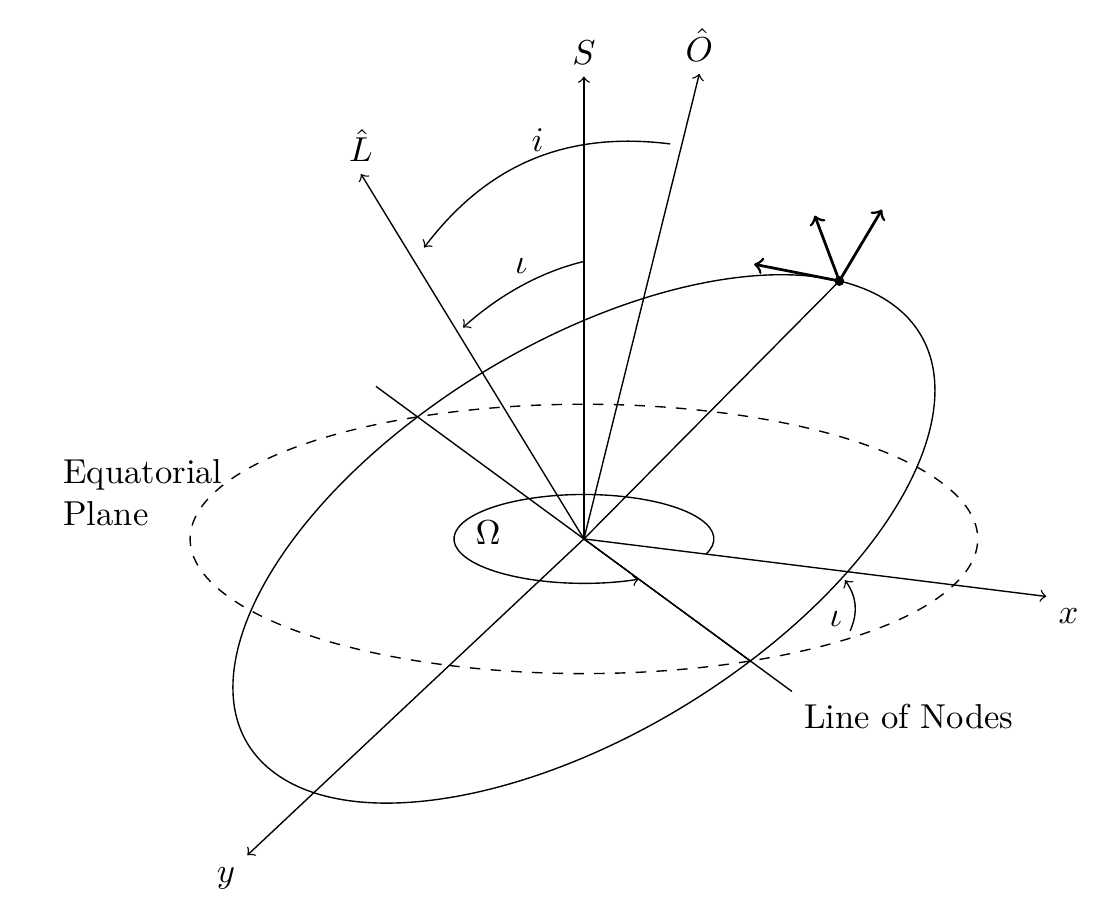}
	\caption{Schematics of the orbital quantities used in this work. The direction of the BH spin axis is $S$ and the observer lies in direction $\hat{O}$. The orbital plane of the pulsar is inclined at an angle $\iota$ with respect to the BH axis and the angle between the observer and the orbital angular momentum $\hat{L}$ is given by $i$. The longitude of ascending mode is given by $\Omega$. }
	\label{fig:orbitalelements}
\end{figure}
As discussed, in Kerr geometry $E,L_z$ and $Q$ of an MSP are conserved along its orbit. For astronomical purposes it is useful to map these constants to orbital parameters. In this way we can specify the sort of Keplerian orbit that we want to describe. In particular, the $E,L_z, Q$ can be mapped to the orbital elements $\mathcal{P}, e, \iota$, the semi-latus rectum, eccentricity and inclination respectively. In the gravitational strong field the semi-latus rectum and the eccentricity are effectively a re-parameterisation in turn of the apsidal approaches (periapsis distance $r_p$ and apoapsis $r_a$),
\begin{eqnarray}
r_p = \frac{\mathcal{P}}{1+e} \, \, \, ; \, \, \, r_a = \frac{\mathcal{P}}{1-e} \ ,
\end{eqnarray}
whilst in the weak field limit $\mathcal{P}, e,$ correspond exactly to the usual Keplerian notion of orbital parameters. The inclination angle $\iota$ is defined \citep{Glampedakis2002}
\begin{eqnarray}
\cos \iota = \frac{L_z}{\sqrt{Q+L_z}} \ .
\end{eqnarray}
We also define $i$ - as distinct from $\iota$- as the inclination of the orbit with respect to  the observer, 
\begin{eqnarray}
\cos i = \hat{\boldsymbol{L}}_z \cdot \hat{\boldsymbol{O}}
\label{eq:inc}
\end{eqnarray}
and $\hat{\boldsymbol{O}}$ is the line of sight unit vector between the observer and the BH. The observer is at some distant radius  with polar and azimuthal coordinates $\Theta, \Phi$ respectively. The relevant geometry is described in Fig \ref{fig:orbitalelements}. The mapping between conserved quantities and orbital parameters is defined in \citep{Schmidt2002, Barausse2007} as 
\begin{eqnarray}
E=\sqrt{\frac{\kappa \rho+2 \epsilon \sigma-2 D \sqrt{\sigma\left(\sigma \epsilon^{2}+\rho \epsilon \kappa-\eta \kappa^{2}\right)}}{\rho^{2}+4 \eta \sigma}} \ ,
\end{eqnarray}
\begin{eqnarray}
L_{z}=-\frac{g_{1} E}{h_{1}}+\frac{D}{h_{1}} \sqrt{g_{1}^{2} E^{2}+\left(f_{1} E^{2}-\text{d}_{1}\right) h_{1}} \ ,
\end{eqnarray}
\begin{eqnarray}
Q=z_{-}\left[a^{2}\left(1-E^{2}\right)+\frac{L_{z}^{2}}{1-z_{-}}\right] \ ,
\end{eqnarray}
where
\begin{eqnarray}
z_{-} = \sin^2 \iota
\end{eqnarray}
and $D = \pm 1$ denotes prograde and retrograde orbits.  In turn, the functions are defined as,
\begin{eqnarray}
f(r) \equiv r^{4}+a^{2}\left[r(r+2)+z_{-} \Delta\right] ' ,
\end{eqnarray}
\begin{eqnarray}
g(r) \equiv 2 a r \ ,
\end{eqnarray}
\begin{eqnarray}
h(r) \equiv r(r-2)+\frac{\Delta z_{-}}{1-z_{-}} \ ,
\end{eqnarray}
\begin{eqnarray}
\text{d}(r) \equiv \Delta \left(r^{2}+a^{2} z_{-}\right) \ ,
\end{eqnarray}
and for the eccentric orbits considered in this work,
\begin{eqnarray}
(f_1, g_1, h_1, d_1) = f(r_p), g(r_p), h(r_p), \text{d}(r_p) \ , \\
(f_2, g_2, h_2, d_2) = f(r_a), g(r_a), h(r_a), \text{d}(r_a) \  ,
\end{eqnarray}
and
\begin{eqnarray}
\kappa & \equiv \text{d}_{1} h_{2}-\text{d}_{2} h_{1} \ , \\
\varepsilon & \equiv \text{d}_{1} g_{2}-\text{d}_{2} g_{1} \ ,\\
\rho & \equiv f_{1} h_{2}-f_{2} h_{1}  \ ,\\ 
\eta & \equiv f_{1} g_{2}-f_{2} g_{1} \ , \\
\sigma & \equiv g_{1} h_{2}-g_{2} h_{1} \ .
\end{eqnarray}
An MSP with particular orbital parameters $\mathcal{P}, e$ will have a corresponding semi-major axis $a_* = \mathcal{P}/(1-e^2)$. As a shorthand we sometimes refer not to the system semi-major axis, but instead to the orbital period $P$ where the two are related via Kepler's 3rd law, $P^2 \propto a_*^3$. We use this only as a convenient re-parameterisation with the understanding that this relation only hold exactly in the weak-field, zero-spin limit and that generally the orbital period will be different due to contributions from BH spin. 

\subsection{Radio beam propagation}
\label{section:raytracing}

We describe the propagation of the pulsar radio signal to the observer in terms of geometrical optics. In this case
the general vacuum Hamiltonian is,
\begin{eqnarray}
H(x^\mu,k_\mu) = \frac{1}{2} g^{\mu \nu} k_{\mu} k_{\nu }   = 0\ , 
\end{eqnarray} 
and the corresponding equations of motion, Hamilton's equations are 
\begin{eqnarray}
\dot{x^{\mu}} = \frac{\partial H}{\partial k_{\mu}} \, , \hspace*{0.2cm} \dot{k}_{\mu} = -\frac{\partial H}{\partial x^{\mu}} \ , 
\end{eqnarray}
where $g_{\mu \nu}$ is the spacetime metric, $k^\mu$ the contravariant 4-momenta, $x^{\mu}$ the spacetime coordinates, and an overdot denotes the derivative with respect to some affine parameter. For a Kerr spacetime and the associated conserved quantities the equations of motion reduce to a problem of quadratures whereby we have four ordinary differential equations $(\dot{t}, \dot{r}, \dot{\theta}, \dot{\phi})$  and four associated constants of motion $E,$ $L_z$, $Q_p$, and $H$.  The system of equations is integrable. It follows that the complete set of equations of motion is given, via Hamilton's equations, as,
\begin{align}
\dot{t} &= E + \frac{2r(r^2 +a^2)E - 2arL_z}{\Sigma \Delta}  \  ;  \label{eq:tdot} \\ 
\dot{r} &= \frac{p_r \Delta}{\Sigma}  \ ;  \\ 
\dot{\theta} &= \frac{p_{\theta}}{\Sigma} \ ;    \\ 
\dot{\phi} &= \frac{2arE + (\Sigma - 2r)L_z\csc^2\theta}{\Sigma \Delta} \ ;   \\ 
\dot{k}_{\theta} &= \frac{1}{2 \Sigma} \left[
-2a^2 E^2 \sin \theta \cos \theta + 2L_z^2 \cot \theta \csc^2 \theta \right] \ ;  \\ 
\dot{k}_r &= \frac{1}{\Sigma \Delta} \biggl[-\kappa(r-1) +2r(r^2+a^2)E^2 - 2aEL\biggl]- \frac{2 p_r^2(r-1)}{\Sigma} 
\label{eq:pr}
\end{align}
where $\kappa = p_{\theta}^2 +  E^2a^2 \sin^2 \theta + L_z^2 \csc ^2 \theta $. The two integration constants, $E$ (energy at infinity) and $L_z$ (the azimuthal component of the angular momentum at infinity), 
can be determined by the initial conditions using the following relations:  
\begin{eqnarray}
E^2 = (\Sigma - 2r) \left(\frac{\dot{r}^2}{\Delta} + \dot{\theta}^2 
+ \right) + \Delta \dot{\phi}^2 \sin^2 \theta   \ ;   \\ 
L_z = \frac{(\Sigma \Delta \dot{\phi}-2arE)\sin^2\theta}{\Sigma - 2r}   \ . 
\end{eqnarray}  
We also integrate this system of equations numerically so as to determine the evolution of the photon ray subject to all general relativistic effects. 

\section{Comparison with Post-Keplerian approximation}
\label{sec:PK}
Given that we can now describe the relativistic orbital dynamics of the pulsar and the propagation of light between pulsar and observer, we are now in a position to investigate the pulsar timing delays for a MSP-E/IMRB. We can compare the fully general relativistic solution with the timing delays given by the PK approximation. For a given quantity $X$ we can determine four `levels' of solution
\begin{enumerate}
	\item $X_{\text{PK}, i}$ - solution given by the $i$-th order PK approximation
	\item $X_{\text{GEO}, a=0}$ - GR solution for zero BH spin ($a=0$) and zero MSP spin (i.e. Schwarzschild geodesic)
	\item $X_{\text{GEO}, a \ne 0}$ - GR solution for non-zero BH spin  and zero MSP spin (i.e. Kerr geodesic)
	\item $X_{\rm MPD}$ - GR solution for spinning BH and spinning MSP (i.e. full MPD solution)
\end{enumerate}
We investigate the difference between each of these levels of solution, defining the respective differences as, 
\begin{eqnarray}
\delta_{\alpha,i} (X) = X_{\text{GEO}, a=0} - X_{\text{PK}, i}
\end{eqnarray}
\begin{eqnarray}
\delta_{\beta}(X) = X_{\text{GEO}, a \ne 0} - X_{\text{GEO}, a=0} 
\end{eqnarray}
\begin{eqnarray}
\delta_{\gamma} (X)= X_{\rm MPD} - X_{\text{GEO}, a \ne 0}
\end{eqnarray}
It is worth taking a moment to consider how we compare different quantities in different spacetimes. For $\delta_{\beta}(X)$ and $\delta_{\gamma} (X)$, the comparison is straightforward; in all cases we are working in Boyer-Lindquist coordinates, with the Schwarzschild coordinates of the Schwarzschild metric simple being a special ($a=0$) case. However, solutions derived within the post-Keplerian framework are derived from the PK metric which typically uses a different coordinate system. In particular, by writing the Schwarzschild metric in harmonic coordinates (such that $g^{\mu \nu} {\Gamma^{\alpha}}_{\mu \nu} = 0 $) where the radial harmonic coordinates $r_h$ is related to the Boyer Lindquist coordinate as,
\begin{eqnarray}
	r = r_h + 1
	\label{eq:harmonic}
\end{eqnarray}
and by taking the weak field (large $r_h$) limit, the metric components become,
\begin{eqnarray} 
	g_{00} =-1+2 / r_{\mathrm{h}}-\frac{1}{2}\left(2 / r_{\mathrm{h}}\right)^{2}+\cdots  
\end{eqnarray}
\begin{eqnarray}
g_{j	 k} &=\left(1+2 / r_{\mathrm{h}}\right) \delta_{j k}+\frac{1}{4}\left(2 / r_{\mathrm{h}}\right)^{2}\left(\delta_{j k}+n_{j} n_{k}\right)+\cdots
\end{eqnarray}
where $n^j = x^j / r_h$ and $n_j = \delta_{jk} n^k$ \citep{poisson2014gravity}. At first order the metric interval is then
\begin{eqnarray}
d s^{2}=-\left(1-\frac{2}{r_h}\right) d t^{2}+\left(1+\frac{2}{r_h}\right)(d \bar{\rho})^{2}
\label{eq:PKmetric}
\end{eqnarray}
where $(d \bar{\rho})^{2} = dx^2 + dy^2 + dz^2$.This is the first order PK metric and is the one typically used \citep[e.g.][]{Blandford1976} for calculating the pulsar timing delays. Explicitly, the second order metric solution is:
\begin{align}
	ds^2 & =  -\left(1-\frac{2}{r_h} + \frac{2}{r_h^2}\right) d t^{2}+\left(1+\frac{2}{r_h} + \frac{1}{r_h^2}\right)(d \bar{\rho})^{2} \\
	& + \frac{1}{r_h^2} n_j n_k dx^j dx^k \ .
	\label{eq:PKmetric2}
\end{align}

\noindent The primary search area for MSP-E/IMRBs is the Galactic centre. Although no MSPs have yet been detected in this region previous surveys can be shown to be insensitive to the population \citep{Rajwade2017} and so the development of advanced radio facilities with increased sensitivities such as the SKA or the dedicated search methods of NASA DSN, for example - along with the development of appropriate relativistic search algorithms -should render this population detectable. Subsequently,  to use an MSP in the Galactic centre as a precision probe of strong-field GR \citep[e.g. ][]{Liu2012, Psaltis2016} it is important to time a pulsar that is solely affected by the gravitational field of the central BH. However there do exist astrophysical perturbations to the pulsar orbit e.g. gravitational perturbations due to the presence of other masses \citep{Merritt2010}. For orbits with periods $ \lesssim 0.3$ year external perturbations are expected to be negligible \citep{Liu2012}. Observations close to periapsis of some longer, sufficiently eccentric orbits may also be viable. Consequently in this work we consider eccentric systems with appropriately short orbital periods $ \lesssim 0.3$ year. We take the mass of the Galactic centre BH to be $M \sim 4.3 \times 10^6 M_{\odot}$ \citep{Gillessen2009}. The PK timing delays that we investigate are,
\begin{enumerate}
	\item Einstein delay, $\Delta_E$ - due to gravitational and relativistic time dilation of the MSP.
	\item Roemer delay, $\Delta_R$ - due to the orbital motion of the MSP
	\item Shapiro delay, $\Delta_S$ - due to the gravitational time dilation of light
\end{enumerate}
We now explore each of these pulsar timing delays in turn

\subsection{Einstein Delay, $\Delta_E$}
The Einstein delay describes the Doppler shift and gravitational time delay (redshift), or  the difference between the MSP proper time and the time measured by a distant observer,
\begin{eqnarray}
\Delta_{\rm E} = t - \tau
\end{eqnarray}
Determining the Einstein delay from the numerical relativistic solutions is straightforward, since at each integration time step we have the MSP proper time and the observer coordinate time by solving Eq. \ref{eq:MPD2}. Within the PK approximation, one can derive an analytical expression for the Einstein delay \citep{Blandford1976,Zhang2017} as,
\begin{eqnarray}
\Delta_E = \frac{P e}{\pi a_{*}} (\sin E' - \sin E'_0) + \frac{3}{2}\frac{t}{a_{*}}
\label{eq:ED}
\end{eqnarray}
where $a_*$ is the semi major axis (as distinct from spin parameter $a$) and $E'$ the eccentric anomaly, an angular parameter given by
\begin{eqnarray}
\cos E' (t) = \frac{1}{e} \left(1 - \frac{r_h(t)}{a_*}\right)
\end{eqnarray}
Evidentially the Einstein delay is then composed of an oscillatory part, of magnitude $\gamma = P e / \pi a_*$ and a linear term. Rather than computing the analytical evolution of $\Delta_{\rm E}$ by e.g. solving the Kepler equation for $\mathcal{E}$, we employ a different approach. From Eq. \ref{eq:PKmetric} it follows that,
\begin{eqnarray}
\left(\frac{dt}{d\tau}\right)^2 = \frac{1 + \left(1+\frac{2}{r_h}\right) v^2 }{1-\frac{2}{r_h}}
\label{eq:einstein_differential}
\end{eqnarray}
where $v^2 =|d^2 \bar{\rho} / d \tau^2 |$. At second order,
\begin{eqnarray}
\left(\frac{dt}{d\tau}\right)^2 = \frac{1 + \frac{1}{r_h^2} n_j n_k \frac{dx^j}{d\tau} \frac{dx^k}{d \tau}  + \left(1+\frac{2}{r_h} + \frac{1}{r_h^2}\right) v^2 }{1-\frac{2}{r_h} + \frac{2}{r_h^2}}
\label{eq:einstein_differential2}
\end{eqnarray}
We can express the squared velocity magnitude in terms of BL coordinates by the straightforward coordinate transformation of a general contravariant vector,
\begin{eqnarray}
Y^{a} = \frac{\partial x^a}{\partial x^b} Y^b
\end{eqnarray}
Explicitly, in conjunction with Eq. \ref{eq:harmonic}, the fact that $dr_h/dr = 1$ and the transformation between (spherical polar) harmonic and Cartesian coordinates then,
\begin{align}
v^2 &= v_x^2 + v_y^2 + v_z^2 \\
 =& (\sin \theta \cos \phi \, u^r + r_h \cos \theta \cos \phi \, u^{\theta} - r_h \sin \theta \sin \phi \, u^{\phi})^2 \nonumber \\ 
 &+ (\sin \theta \sin \phi \, u^r + r_h \cos \theta \sin \phi \, u^{\theta} - r_h \sin \theta \cos \phi \, u^{\phi})^2 \nonumber \\
 &+ (\cos \theta \, u^r - r_h \sin \theta \, u^{\theta})^2
\end{align}
where the $u^{\alpha}$ terms are the components of the MSP 4-velocity as given by Eq. \ref{eq:MPD2}. This gives us a differential equation for $t$ as determined by the PK metric. It is equivalent to the 0-th term of the relativistic solution described by Eq. \ref{eq:MPD2}. As we integrate the complete relativistic solution numerically, we can then at each integration timestep also compute $v$ and $r_h$ and solve Eq. \ref{eq:einstein_differential}. This then gives us $t$ (and hence $\Delta_{\rm E}$) as given by the PK solution for an MSP which has the same spatial orbital trajectory as described by the full relativistic solution.
\begin{figure}
	\includegraphics[width=\columnwidth]{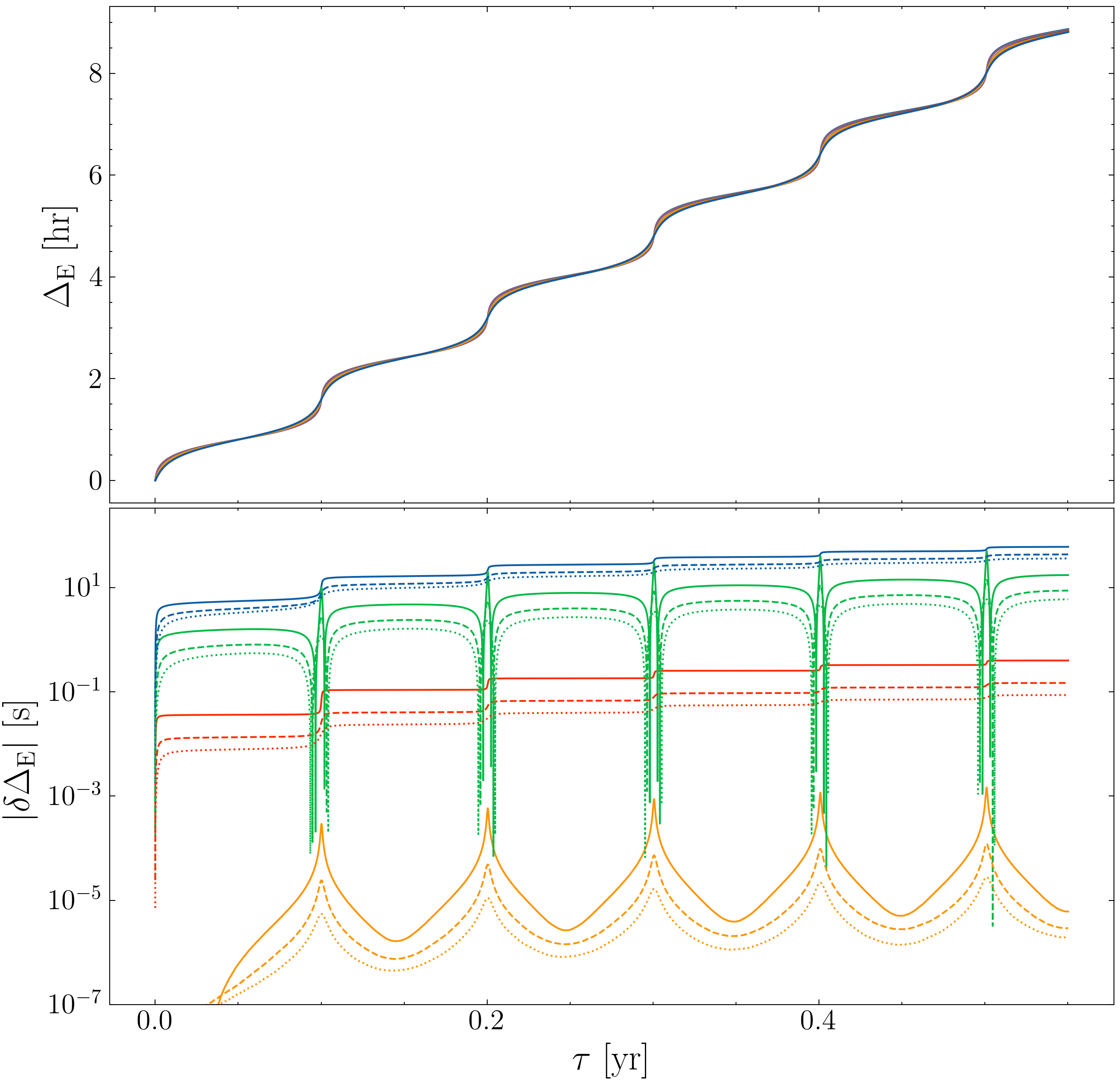}
	\caption{\textbf{Top panel:} Einstein delay given by the 4 solutions (1 PK + 3 GR) from an MSP orbiting the Galactic centre BH in the equatorial plane at a range of difference eccentricities. The BH mass is set as $4.31 \times 10^6 M_{\odot}$, with spin parameter $a+0.6$. The delay accumulates up to $2$ hrs after one orbital period, with rapid variations as the pulsar goes through periapsis.\textit{ Bottom panel:} Differences between the respective solutions $\delta_{\alpha,1} (\Delta_{\rm E})$ (blue),$\delta_{\alpha,2} (\Delta_{\rm E})$ (red), $\delta_{\beta} (\Delta_{\rm E})$ (green), $\delta_{\gamma} (\Delta_{\rm E})$ (orange) at eccentricities $e=0.9,0.8,0.7$ (solid, dashed, dotted lines respectively) }
	\label{fig:einsteindelay}
\end{figure}
The Einstein delay given by the respective solutions for a Galactic centre MSP with  orbital period $P=0.1$ years, in the equatorial plane is presented in Fig. \ref{fig:einsteindelay} at eccentricities $e=0,7,0.8,0.9$. The evolution of the Einstein delay for all the orbits considered has general oscillatory contribution, illustrated by the first term of Eq. \ref{eq:ED} and an additional linear term. Rapid variations are observed as the MSP passes through periapsis and is moving at its fastest. After a single orbital period the Einstein delay has accumulated to $\sim 2$ hrs for all the solutions and all the eccentricities. Variations in the Einstein delay induced by relativistic effects are seen to occur at smaller scales: at the first level, $\delta_{\alpha,1} (\Delta_{\rm E})$ is of the order $10^1$ seconds, whilst $\delta_{\alpha,2} (\Delta_{\rm E}) \sim 10^{-2}-10^{-1}$ \textbf{seconds}, with greater magnitude errors for more eccentric orbits which move faster at periapsis and probe a stronger gravitational field. The variations induced by the BH spin for $a=+0.6$ , $\delta_{\beta} (\Delta_{\rm E})$ introduce additional variations in the timing delay on the order of 0.1 -1 seconds, again with more eccentric orbits exhibiting larger magnitude variations. The influence of `switching on' the MSP spin and the associated couplings is particularly interesting. Although subdominant to the $\delta_{\alpha, \beta}$ variations, the MSP spin introduces periodic contributions to the Einstein delay. As before, these corrections peak in magnitude at periapsis and for $e=0.9$, $\delta_{\gamma} (\Delta_{\rm E}) \sim 1$ ms, whilst for $e=0.7$ the maximal variations are of magnitude $\sim 30$ $\mu$s. Radio facilities such as FAST or SKA are able to determine pulse ToA to precisions $< 100 $ ns \citep{Hobbs2014} \citep{Stappers2018}, especially for stable MSPs which are free of glitches. Consequently, small variations in the timing residuals $\mathcal{O}$  ($\mu$s - ms) will be both detectable and important for a consistent, long-term timing solution.

\subsection{Propagation Delays: Shapiro and Roemer, $\Delta_S$ , $\Delta_R$}
In PK timing models for pulsar binaries, the photon propagation time between observer and source is the sum of the Roemer delay ($\Delta_{\rm R}$, i.e. time delay due to the finite speed of light propagating in a flat spacetime) and the Shapiro delay ($\Delta_{\rm S}$, time delay due to first order curvature corrections). This can be seen by decomposing the spacetime geodesic in terms of a Minkowskian (0th order) background plus a higher-order curvature perturbation \citep[see e.g.][]{Carroll2004} i.e. 
\begin{eqnarray}
x^{\mu} = x^{(0)\mu} +x^{(1)\mu} + \boldsymbol{x^{(2)} \mu} \,
\end{eqnarray}
The background and perturbation tangent vectors are,
\begin{eqnarray}
k^{\mu} \equiv \frac{dx^{(0)\mu}}{ d \lambda} \, \, ; \, \,  l^{\mu} \equiv \frac{dx^{(1)\mu}}{ d \lambda} \, \, ; \, \,  q^{\mu} \equiv \frac{dx^{(2)\mu}}{ d \lambda}
\end{eqnarray}
By considering the perturbed geodesic equation, it can be shown that,
\begin{eqnarray}
\frac{d}{d \lambda} l^{\mu} = -\prescript{}{(1)}{\Gamma^{\mu}}_{\rho \sigma} k^{\rho} k^{\sigma}
\label{eq:dl}
\end{eqnarray}
\begin{eqnarray}
\frac{d}{d \lambda} q^{\mu} = -\prescript{}{(2)}{\Gamma^{\mu}}_{\rho \sigma} k^{\rho} k^{\sigma} - \prescript{}{(1)}{\Gamma^{\mu}}_{\rho \sigma}(k^{\rho}l^{\sigma} + k^{\sigma}l^{\rho})
\label{eq:dq}
\end{eqnarray}
where $\prescript{}{(i)}{\Gamma^{\mu}}_{\rho \sigma}$ denotes the $i$-th order terms of the Christoffel connection. It then follows that the photon propagation time is,
\begin{align}
\Delta_{\rm prop}&= \int \frac{dx^0}{d \lambda} d \lambda \nonumber \\
&= \int k^0 d\lambda + \int l^0 d\lambda+ \int q^0 d\lambda \nonumber \\ 
&= \Delta_{\rm R} + \Delta_{\rm S} + \Delta_{\rm S}^{(2)}
\label{eq:FlightTime}
\end{align}
where $\Delta_{\rm S}^{(2)}$ labels the higher (second) order curvature time delay. Within the PK framework, one can derive analytical expressions for the evolution of $\Delta_{\rm R}$, $\Delta_{\rm S}$ in terms of the system orbital parameters as \citep{Shapiro1964,Blandford1976,Damour1986,Zhang2017},
\begin{eqnarray}
 \Delta_{\mathrm{R}}=\tilde{\alpha}\left(\cos E^{\prime}-e \right)+\tilde{\beta} \sin E^{\prime}
 \label{eq:roemer:}
\end{eqnarray}
\begin{eqnarray}
\Delta_{\mathrm{S}}=2 \ln \left[\frac{1+e \cos f}{1-\sin i \sin \left(\omega+f\right)}\right]
\label{eq:shapiro}
\end{eqnarray}
where $i$ is the inclination with respect to the observer, defined via Eq. \ref{eq:inc}, $\omega$ is the angle of periapsis and $f$ the true anomaly, with
\begin{eqnarray}
\tilde{\alpha}=a_{\star}  \sin i \sin \omega
\end{eqnarray}
\begin{eqnarray}
\tilde{\beta}=\left(1-e^{2}\right)^{1 / 2} a_{\star}  \sin i \cos \omega
\end{eqnarray}
Similar to deriving the Einstein delay in the previous section, we do not use these expressions explicitly but instead proceed via an equivalent route to express the time delays in terms of coordinates rather than orbital parameters. The propagation time between two points $\boldsymbol{r}_1, \boldsymbol{r}_2$
can be calculated by first determining the temporal components of the tangent vectors by integrating Eqs. \ref{eq:dl}, \ref{eq:dq} 
\begin{eqnarray}
l^{0} = -2k \Phi
\label{eq:dl2}
\end{eqnarray}
\begin{eqnarray}
q^0 =2k\Phi^2
\label{eq:dq2}
\end{eqnarray}
 and so the propagation time is given from Eq. \ref{eq:FlightTime} as,
\begin{align}
\Delta_{\rm prop} &= \int ds - 2 \int \Phi ds + 2 \int \Phi^2 ds \nonumber \\
&=|\boldsymbol{r}_2 - \boldsymbol{r}_1| + 2 \int \frac{1}{ | \boldsymbol{r}|} ds+ 2 \int \frac{1}{ | \boldsymbol{r}|^2} ds
\end{align}
where $ds$ denotes the unperturbed background path.
Considering the second integral term (i.e. the Shapiro delay,  $\Delta_{\rm S}$), we can parameterise in terms of $t$ such that,
\begin{eqnarray}
\boldsymbol{r} = \boldsymbol{r}_1 + \boldsymbol{\hat{u}} t
\end{eqnarray}
where $\boldsymbol{\hat{u}}$ is the unit vector,
\begin{eqnarray}
\boldsymbol{\hat{u}} = \frac{\boldsymbol{r}_2 - \boldsymbol{r}_1}{|\boldsymbol{r}_2 - \boldsymbol{r}_1|}
\end{eqnarray}
and $t \in [0,|\boldsymbol{r_2} - \boldsymbol{r_1}|]$. With this parameterisation, $ds = |\boldsymbol{\hat{u}}| dt$ and so,
\begin{align}
\Delta_{\rm S} &= \int \frac{1}{ |\boldsymbol{r}_1 + \boldsymbol{\hat{u}} t|} dt \nonumber \\
&= \int \frac{1}{ \sqrt{|\boldsymbol{r}_1|^2 + t^2 + 2t \boldsymbol{r}_{1} \cdot \boldsymbol{\hat{u}}    }} dt
\end{align}
Performing the integral gives,
\begin{eqnarray}
\Delta_{\rm S} = \left[ \ln \left(t + \boldsymbol{r}_1 \cdot \boldsymbol{\hat{u}} + |\boldsymbol{r}_1 + \boldsymbol{\hat{u}} t|  \right) \right]_{t=0}^{t = |\boldsymbol{r}_2 - \boldsymbol{r}_1|} \ ,
\end{eqnarray}
and so,
\begin{eqnarray}
\Delta_{\rm S} = \ln \left( \frac{|\boldsymbol{r}_2 - \boldsymbol{r}_1| + \boldsymbol{r}_1 \cdot \boldsymbol{\hat{u}} + |\boldsymbol{r}_2|}{\boldsymbol{r}_1\cdot \boldsymbol{\hat{u}} + |\boldsymbol{r}_1|} \right)
\end{eqnarray}
Using the same parametrization for the quadratic term, it follows that
\begin{align}
\Delta_{\mathrm{S}}^{(2)} &= \left[ \frac{1}{\zeta} \arctan \left(\frac{t+\boldsymbol{r}_{1} \cdot \boldsymbol{\hat{u}}}{\zeta}\right)\right]_{t=0}^{t = |\boldsymbol{r}_2 - \boldsymbol{r}_1|} \\
&= \frac{1}{\zeta} \arctan \left(\frac{|\boldsymbol{r}_2 - \boldsymbol{r}_1|+\boldsymbol{r}_{1} \cdot \boldsymbol{\hat{u}}}{\zeta}\right) - \frac{1}{\zeta} \arctan \left(\frac{\boldsymbol{r}_{1} \cdot \boldsymbol{\hat{u}}}{\zeta}\right) 
\end{align}
with $\zeta = \sqrt{|\boldsymbol{r}_1|^2 -(\boldsymbol{r}_{1} \cdot \boldsymbol{\hat{u}})^2 }$
The propagation time along a straight line between two points in the PK framework at second order is then,
\begin{align}
\Delta_{\rm prop} &=|\vec{r}_2 - \vec{r}_1| + 2 \ln \left( \frac{|\vec{r}_2 - \vec{r}_1| + \vec{r}_1 \cdot \hat{u} + |\vec{r}_2|}{\vec{r}_1\cdot \hat{u} + |\vec{r}_1|} \right) \\ &= \frac{1}{\zeta} \arctan \left(\frac{|\boldsymbol{r}_2 - \boldsymbol{r}_1|+\boldsymbol{r}_{1} \cdot \boldsymbol{\hat{u}}}{\zeta}\right) - \frac{1}{\zeta} \arctan \left(\frac{\boldsymbol{r}_{1} \cdot \boldsymbol{\hat{u}}}{\zeta}\right)  \ ,
\label{eq:tPK}
\end{align}
where the first term corresponds to the Roemer delay and the second and third terms to the Shapiro delay at first and second order respectively. A linear decomposition of the propagation time in this way is not generally possible for the relativistic solutions, instead being found by solving Eq. \ref{eq:tdot}. Therefore going forward, we consider the Roemer and Shapiro terms together in terms of the general propagation delay $\Delta_{\rm prop}$. \newline 

\begin{figure*}
	\includegraphics[width=\linewidth
	]{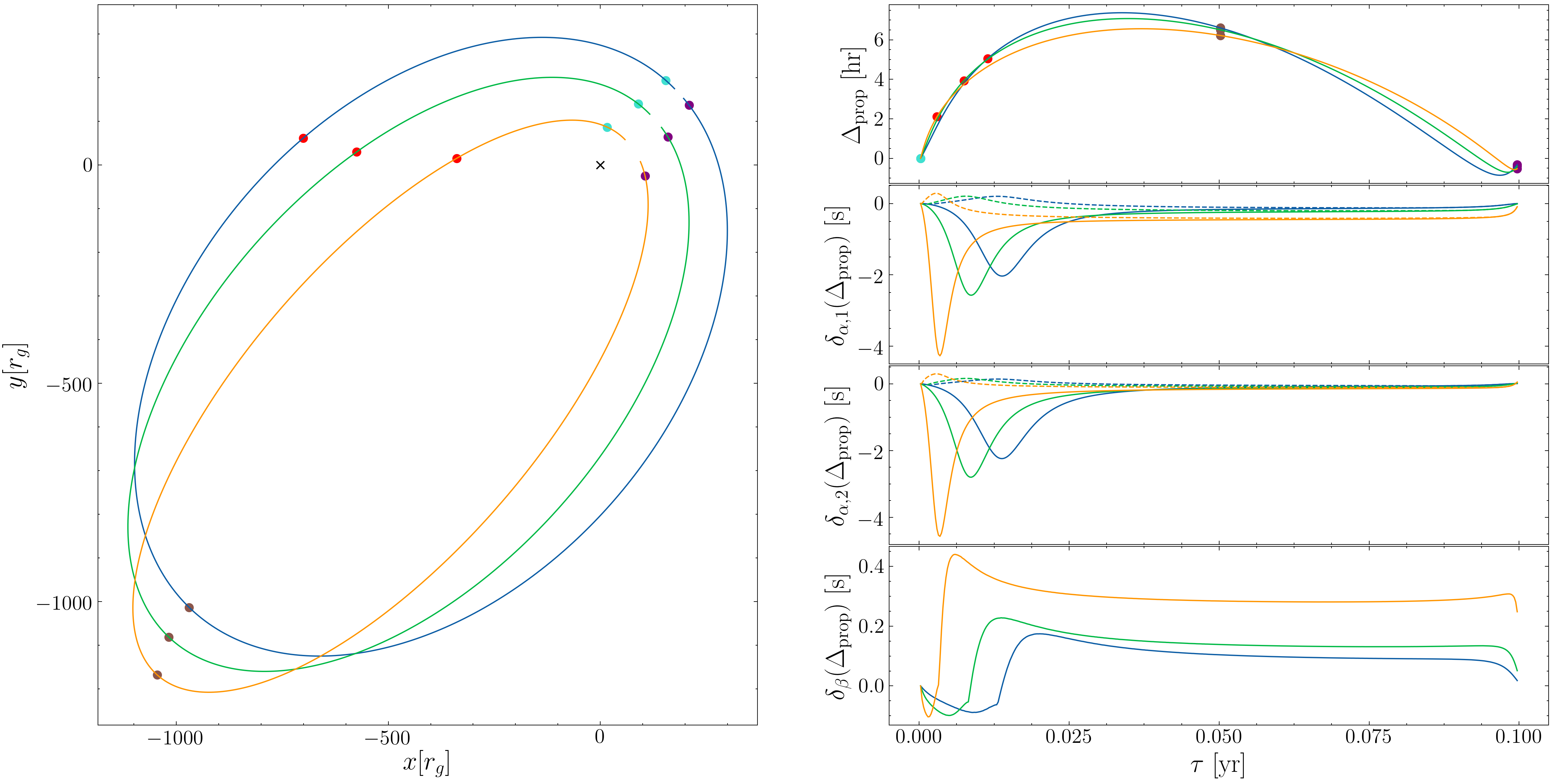}
	\caption{(\textit{Left panel:}) Orbital trajectories of a Galactic centre MSP with $a_*=825 r_g$, $i=76^\circ$, $\omega = \Omega = \pi/4$ and eccentricities $e=0.7,0.8,0.9$ (blue,green,orange lines respectively). The scatter points are included as a reference for the right panels and label the initial, final,middle and superior conjunction sample points (turquoise,purple,brown,red respectively) The black cross labels the BH singularity. (\textit{Right panels:}) The propagation timing delay for each of the eccentric orbits, along with the errors between the PK and relativistic solutions ($\delta_{\alpha, i}(\Delta_{\rm prop})$,$\delta_{\beta}(\Delta_{\rm prop})$). For the middle two panels the dashed lines show the residual when the PK lensing correction is included (see also Fig. \ref{fig:flightdelay_lensing} ). Significant timing residuals are present between both $\delta_{\alpha, i}(\Delta_{\rm prop})$,$\delta_{\beta}(\Delta_{\rm prop})$ relative to typical radio pulsar timing precision.}
	\label{fig:flightdelay}
\end{figure*}
\noindent To explore the propagation delay according to the PK and relativistic solutions, we take a discrete sample of position points of an eccentric orbit. The orbit is calculated by solving the Kerr orbital geodesic equations numerically. The propagation delay between each sample point (i.e. the pulsar emitter) and the image plane of a distant observer (with Boyer Lindquist coordinates $R = 10^4, \Theta = \pi/2, \Phi = 0$) can be calculated via the PK solution ( Eq. \ref{eq:tPK}) and the relativistic solutions (Eq. \ref{eq:tdot}.) Obviously, since the MSP spin does not influence the photon propagation, $\delta_{\gamma} (\Delta_{\rm prop}) = 0$ (however, the MSP spin does generally influence the orbital evolution of the pulsar which in turn will influence the evolution of $\Delta_{\rm R}$. We return to this point later.) We consider the orbit of a Galactic centre MSP with semi-major axis $a_* = 825 r_g$, $\iota = $, $\omega = \Omega = \pi/4$ and $\iota=20^\circ$ at eccentricities $e=0.7,0.8,0.9$. The spin parameter is $a=+0.6$.  Since the observers image plane is in the positive $x$-direction, this corresponds to an inclination with respect to the observer of $i=76^\circ$. The propagation delay, along with the associated orbital trajectory, for this system is shown in Fig. \ref{fig:flightdelay}. Since generally pulsar timing can only detect variations in the light travel time, rather than the absolute travel time itself, we normalise $\Delta_{\rm prop}$ with respect to its initial value. The propagation delay evolves on the scale of hours over the course of the orbit, with more eccentric orbits having a slightly greater magnitude of variation. The error in the PK solutions at first and second order compared to the Schwarzschild relativistic solutions, $\delta_{\alpha, i} (\Delta_{\rm prop})$ evolve over the orbit and reaches an extrema when the pulsar is on the far side of the BH. In this case the photon tray traverses a more strongly curved spacetime and is subject to a greater degree of strong-field time dilation and gravitational lensing, both of which drive the PK solutions away from the relativistic solution. For this orientation, more eccentric orbits have greater extrema than less eccentric orbits since their far side points are at shorter orbital radii. The additional propagation delays induced by the BH spin, $\delta_{\beta} (\Delta_{\rm prop})$, follow a similar general evolution to $\delta_{\alpha} (\Delta_{\rm prop})$, with greatest variations found for eccentric orbits on the far side of the BH. Variations here are of the order $\sim 0.1$ s. Whilst for both $\delta_{\alpha}(\Delta_{\rm prop})$ and $\delta_{\beta} (\Delta_{\rm prop})$ are greatest for far side sample points, the error remains significant over the entire orbit, especially given the high timing precision of pulsar timing ($10-100$ ns). \newline 

\begin{figure}[t]
	\includegraphics[width=\linewidth]{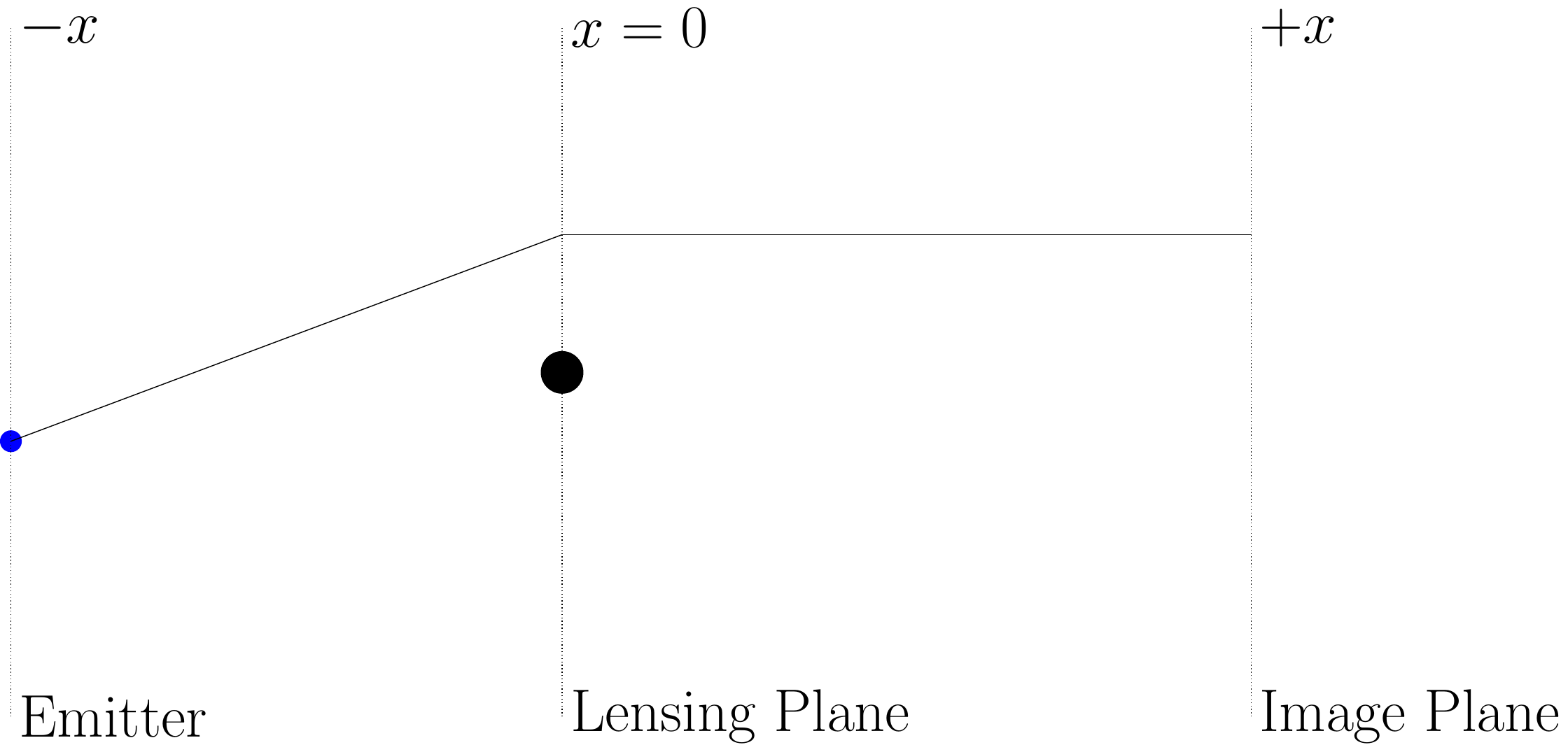}
	\caption{Schematic of weak-field, linear gravitational lensing.}
	\label{fig:lensing}
\end{figure}

\begin{figure}
	\includegraphics[width=\linewidth]{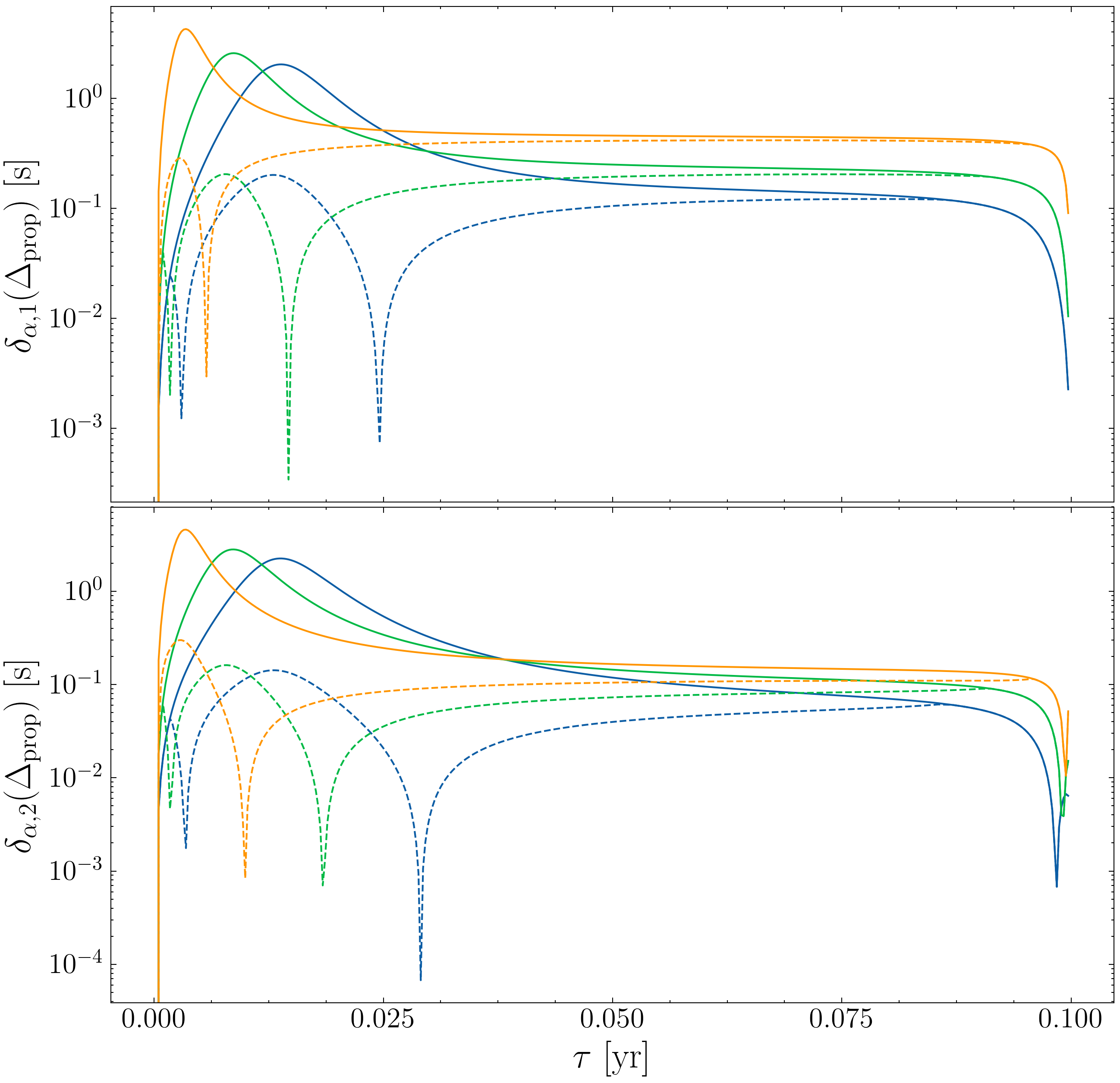}
	\caption{Enhanced view of the two middle right panels of Fig. \ref{fig:flightdelay}. Error between the PK and Schwarzschild propagation delays $\delta_{\alpha,i}(\Delta_{\rm prop})$, for each of the eccentric systems when lensing corrections are included (dashed lines) and without. Lensing significantly improves the solution, but marked residuals remain. The typical magnitude of $\delta_{\alpha,2}(\Delta_{\rm prop})$ are less than $\delta_{\alpha,1}(\Delta_{\rm prop})$, but even in the case where we consider the second order solution with lensing, significant timing discrepancies are present.}
	\label{fig:flightdelay_lensing}
\end{figure}

\begin{figure}
	\includegraphics[width=\columnwidth]{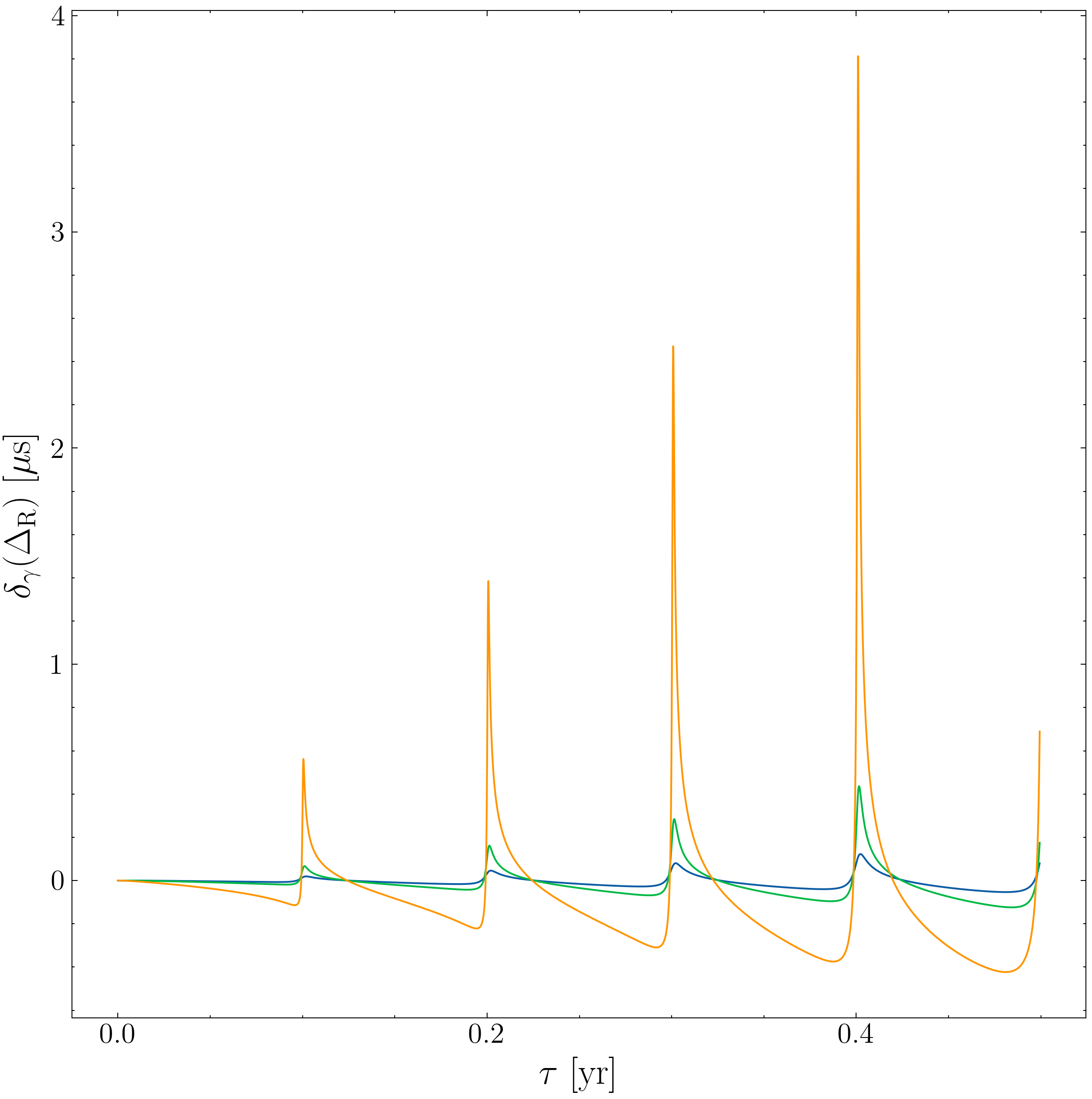}
	\caption{Error in the Roemer delay between the MPD and Kerr solutions. We consider a Galactic centre MSP with $a_*=825 r_g$, $i=76^\circ$, $\omega = \Omega = \pi/4$ and eccentricities $e=0.7,0.8,0.9$ (blue,green,orange lines respectively). Rapid, large magnitude variations are observed as the pulsar goes through periapsis, with the most significant discrepancies - $\mathcal{O}(\mu$s) - for the most eccentric systems which probe the strongest gravitational fields.}
	\label{fig:roemerdelay}
\end{figure}

\noindent Clearly the exact propagation time is going to depend upon the ray path. To first order in the PK solution rays travel in straight lines. Such an approximation may be appropriate for longer orbits and emission on the near side of the BH, but for more compact orbits where the pulsar is on the far side of the BH it is also important to account for the gravitational lensing which is naturally included by the relativistic solution. Lensing introduces two additional timing delays with the PK framework. The first is a geometrical delay $\Delta_{\rm geo}$ due to the different spatial path traversed given by \citet{Rafikov2006},
\begin{eqnarray}
\Delta_{\rm geo} = 2 \left( \frac{\Delta R_{\pm}}{R_E} \right)
\end{eqnarray}
where $R_E$ is the Einstein radius and $\Delta R_{\pm} = |R_{\pm} - R_s|$ for image position $R_{\pm}$ and `flat space position' $R_s$ \citep[see ][ for a further explanation of these terms]{Rafikov2006}. The second delay is due to a correction to the Shapiro delay since the lensed ray traverses a different gravitational potential than the unlensed ray. Indeed, for edge on orbits Eq. \ref{eq:shapiro} diverges when $f = \pi/2$ since this ray path involves going directly through the BH singularity, which is evidently unphysical. At first order within the PK framework, the Shapiro delay is modified due to lensing as \citep{Rafikov2006},
\begin{eqnarray}
\Delta_{\rm S, L} = 2 \ln \left[\frac{a_* (1-e^2)}{\sqrt{r_{||}^2 + R_{\pm}^2} - r_{||}}\right]
\end{eqnarray} 
The net propagation delay the becomes
\begin{eqnarray}
\Delta_{\rm prop} = \Delta_{\rm R}  +\Delta_{\mathrm{S,L}} + \Delta_{\rm geo}
\end{eqnarray}
The linear weak field description of lensing involves decomposing the ray trajectory into two parts. The first is the radial ray which travels from the image plane to the lensing plane. Upon reaching the lensing lane the ray is deflected and travels again in a straight line towards the emitter (see Fig. \ref{fig:lensing}). 
The total propagation time is then the sum of the propagation time along these two linear trajectories. We can therefore the describe the PK time propagation delay taking in account lensing by evaluating Eq. \ref{eq:tPK} along the two relevant trajectories and taking the sum. The effect of the inclusion of PK lensing corrections on $\delta_{\alpha,i} (\Delta_{\rm prop})$is shown in Figs. \ref{fig:flightdelay} and \ref{fig:flightdelay_lensing}. The inclusion of lensing corrections significantly improves the PK solution, with the peak magnitude of $\delta_{\alpha} (\Delta_{\rm prop})$ reducing to $\mathcal{O} (0.1)$ for both the first and second order solutions, with the second order solution generally exhibiting lower magnitudes than the first. However, this error is still far in excess of the timing precisions that can be achieved via pulsar timing. It is also worth noting that whilst lensing is most prominent for points on the far side of the BH with respect to the observer, due to the BH spin the rays are deflected from their Minkowski straight line geodesics even on the near side, if sufficiently close to the BH. \newline
As mentioned previously, the MSP spin will generally couple to the background spacetime and so influence the orbital evolution, driving it away from geodesic motion.  These positional variations will in turn influence the evolution of the Roemer delay. The Roemer delay is generally given as,
\begin{eqnarray}
\Delta_{\mathrm{R}}(\tau) = \boldsymbol{\hat{O}} \cdot \boldsymbol{x}(\tau)
\end{eqnarray}
where $\boldsymbol{\hat{O}}$ is the position unit vector of the observer and $\boldsymbol{x} (\tau)$ the location of the orbiting pulsar. In this way positional variations will imprint on the pulsar timing residuals. The difference in the Roemer delay between a non-spinning and a spinning pulsar, i.e. $\delta_{\gamma} (\Delta_{\mathrm{R}})$, is presented in Fig. \ref{fig:roemerdelay}. 
Generally the MSP spin-curvature coupling manifests in two ways: an additional component perpendicular to the orbital plane \citep[e.g.][]{Singh2014} and also a contribution to the precession of periastron \citep[e.g.][]{Li2019} (which is itself a PK parameter). We consider the same eccentric systems and observer orientation as used previously ($a_* = 825 r_g, i = 76^{\circ}$ etc.). For these orbital parameters, $\delta_{\gamma} (\Delta_{\mathrm{R}})$ varies periodically, with rapid, large magnitude variations as the pulsar goes through periapsis. For the most eccentric systems ($e=0.9$) the extrema of $\delta_{\gamma} (\Delta_{\mathrm{R}})$ are of magnitude a few $\mu$s, whilst for less eccentric systems the magnitude of these variations drops markedly (e.g. $|\delta_{\gamma} (\Delta_{\mathrm{R}})| \sim 0.1 \mu$s for $e=0.8$.) This is on account of more eccentric orbits probing a stronger gravitational field as they pass through periapsis and so the curvature coupling with the MSP spin becomes more significant. Whilst the magnitude of these spin-induced variations are much smaller than the lower order ($\delta_{\alpha, \beta}$) differences there are a few points to consider. Firstly, for the most eccentric orbits these variations are within the timing precision of the next generation of radio facilities. Consequently, in order to use these systems for precision parameter estimation of the central BH such small variation may need to be included in the residuals timing solution. Moreover, whilst we have focused on systems with $P \sim 0.1$ years, for more compact orbits or different observer orientations this spin induced variation will be more pronounced. For stronger gravitational fields - precisely the region in which we wish to test GR - such effects will be more significant. Since the error between the two solutions grows with time upon repeated periapsis passages, systems which are observed over longer timescales may also need to correct for this effect. 

\section{Discussion}
\noindent Estimates based on PK frameworks suggest that via pulsar timing the parameters such as the mass and spin of the BH at the Galactic centre can be determined to a precision $\sim 10^{-5} - 10^{-3}$ \citep[][]{Liu2012,Psaltis2016}. Some authors consider consistent timing campaigns \citep{Liu2012} whilst others \citep{Psaltis2016} consider instead dense observations close to periapsis. Both approaches have merit, however observations close to periapsis seems highly desirable for the purposes of minimising external perturbations \citep{Merritt2010} and maximising the value of the gravitational potential $\epsilon$ which can be investigated. \newline

\noindent However, our results highlight that these regions that are most scientifically attractive (c.f. tests of astrophysics and fundamental physics) are precisely the regions where the accuracy and appropriateness of the PK framework is lacking.  Any pulsar timing model is fundamentally a mapping between the proper time of the pulsar and the time recorded by the observer, $\tau \rightarrow t$. In these regions of high spacetime curvature, the convolution of spin, relativistic and gravitational effects become more prominent and the naive PK framework fails to account for these higher-order effects and the $\tau \rightarrow t$ mapping does not hold. In particular:
\begin{itemize}	
	\item For a $P=0.1$ year orbit at the Galactic centre, the error in the Einstein delay according to the PK framework is of the order $\sim$ 10s of seconds compared to the relativistic (Schwarzschild) solution. This error grows with time; after 5 orbits with eccentricity $e=0.9$, $\delta_{\alpha}(\Delta_{\rm E}) = 55$s. Higher order corrections in the time delay due to the BH spin and the pulsar spin are of the order $10^{-1}$ and $10^{-4}$ seconds respectively (see Fig. \ref{fig:einsteindelay}). The most severe errors are seen at periapsis when the pulsar is moving fastest and the gravitational potential is strongest.
	
	\item The propagation delay for a similar eccentric Galactic centre system, inclined with respect to the observer at $i=76^{\circ}$ varies on the order of hours over the course of a single orbit. The error in the PK solution is $\delta_{\alpha} (\Delta_{\rm prop}) \sim \mathcal{O}(s)$ whilst effects due to the BH spin introduce errors  $\delta_{\alpha} (\Delta_{\rm prop}) \sim \mathcal{O}(0.1 s)$  (Fig. \ref {fig:flightdelay}). For this particular orbital configuration, these errors are most drastic for the more eccentric orbits and the largest magnitude variations are seen when the pulsar is on the far side of the BH; in this case the photon ray must traverse the most strongly curved spacetime.
	
	\item The introduction of first order linear lensing corrections which account for the geometrical time delay $\Delta_{\rm geo}$ and the modified Shapiro delay $\Delta_{\rm S,L}$ dramatically improves the PK propagation solution, reducing $\delta_{\alpha} (\Delta_{\rm prop})$ to $ \sim \mathcal{O}(0.1 s)$  (see Fig. \ref {fig:flightdelay_lensing}) . However, this error is still much greater than typical timing precisions enjoyed by pulsar radio timing.  Consequently higher order corrections will be needed for strong field systems if viewed from a sufficiently `edge on' angle. Even for emission on the near side of the BH, the BH spin will cause an additional lensing effect. 
	
	\item The coupling of the pulsar spin to the background spacetime will cause the orbital dynamics to divert from that of pure geodesic motion. Such a variation on the coordinate variables will manifest in a change in the evolution of the Roemer delay. This effect is small, with $\delta_{\gamma} (\Delta_{\rm R}) \sim 1 \mu$s for a system with $P=0.1$ years and $e=0.9$, observer at $i = 76^{\circ}$, but may still be significant over longer timescales, for more compact systems, or dense observations close to periapsis.

\end{itemize}
Evidently in the gravitational strong-field there is significant discrepancy between the PK and GR solutions across all of the parameters considered in this work. The timing delay according to the PK solution disagrees with the GR solution by at least a few $\mu s$ in all parameters, up to discrepancies as larges as a few $s$ and naturally this error is more severe as relativistic effects become more prominent; i.e. for highly spinning BHs, measurements near periapsis or rays which traverse a spacetime which is highly curved. Consequently, whilst the quoted precisions (e.g. BH mass to precision $10^{-5}$) may be achievable, this work suggests that in EMRB systems which probe the strong-field, higher-order relativistic effects should be included in a complete timing model to accurately determine the prospects of using PSR to test GR. Indeed, the EMRBs discussed in this work are the progenitors of the key LISA target Extreme Mass Ratio Inspirals (EMRIs), where weak-field methods are recognised as being entirely inadequate \citep{Barack2019}. \newline 

\noindent The disagreement between the PK and GR solutions in strong-field regimes motivates the necessity of a consistent, relativistic,frequency-dependent timing model \citep[e.g.][]{Kimpson2019b}. In developing this work, it would be highly desirable to use such a strong-field PSR-timing model to generate an accurate, relativistic, set of mock data. The ToAs generated by this relativistic model can then be used for a consistent covariance analysis \citep[e.g.][]{Liu2012}, fed into a pulsar timing software package \citep[e.g. TEMPO2,][]{TEMPO}, allowing the current pulsar data analysis algorithms to be assessed in the strong-field (c.f. lack of detections in the Galactic Centre) and allow for quantitative estimates on the precision to which the BH parameters can be recovered. We defer this study for future work. In addition, whilst these results are indicative of the appropriateness or otherwise of a PK timing model for MSP-E/IMRBs, for a more definitive work it is necessary to consider a consistent phase-connected solution. We have also not thoroughly explored the whole parameter space instead just considering some typical example system; determining for what sort of set of orbital parameters relativistic effects start to influence the timing solution would also be a worthwhile enterprise. \newline

\noindent Beyond the failure of the PK framework in the strong-field, there also exist additional corrections that will influence the photon ToA and pulse profile, such as spatial and temporal dispersion \citep{Kimpson2019a} precession of the spin axis and relativistic aberration. Being able to accurately model the photon $t-\nu$ signal in relativistic regimes is essential to maximise the scientific return of the detection of pulsar systems \citep[including coincident multimessenger detections, e.g.][]{Kimpson2020GWBurst}, realistically estimate the scientific prospects (e.g. to what precision can we hope to determine the BH mass?) and compare pulsar timing as a GR probe with complementary methods (e.g. Event Horizon telescope, stellar orbits, gravitational wave astronomy). It is worth noting however that the strengths of the PK method lie beyond the timing accuracy; the advantage of PK is their computational simplicity and the straightforward way in which deviations from GR can be incorporated. The numerical methods discussed in this work are naturally much more computationally demanding than the analytical PK solutions, however it is conceptually straightforward to develop the numerical approach beyond the GR solution and include beyond GR effects via the use of some alternative (non-Kerr) metric e.g. the quasi-Kerr metric of \citep{Glampedakis2006} and used in e.g. \cite{Kimpson4}.  \newline

\noindent In addition to correctly modelling strong-field PSR-EMRBs, another important consideration is the best environment to attempt to use PSRs to undertake precision, strong-field GR tests. Whilst much of the cited literature focuses on the Sgr A* BH, owing to its high mass-to-distance ratio, observations of the Galactic centre are made all the more difficult since it is not a clean environment. Material along the line of sight can cause scattering which results in temporal smearing of the pulse profile, whilst  ionized gas can decrease the received flux density. Indeed, since the GC pulsar population is expected to be dominated by MSPs, such environmental effects mean that previous GC searches have been insensitive to the pulsar population \citep{Macquart2015, Rajwade2017}. Perhaps unsurprisingly then, only one pulsar - the magnetar PSR J1745-29 \citep{Kennea2013} - has been detected within 10' of Sgr A*. \newline 

\noindent Whilst the GC remains an important target, potentially more fruitful hunting grounds are globular clusters and dwarf elliptical galaxies. These systems have exceptionally high stellar densities - up to $10^6$ stars per cubic parsec in the central regions \citep{Freire2013}. Due to the effects of mass segregation and dynamical friction, heavy objects like neutron stars are expected to sink to the centre of these stellar clusters. It is observed that per unit mass there are up to $10^3$ more pulsars in globular clusters than in the Galactic disk \citep{Freire2013}. Millisecond pulsars are thought to evolve from low mass X-ray binaries and globular clusters are known hosts of abundant LMXB populations. Consequently MSPs comprise a significant fraction of the globular cluster pulsar population \citep{Camilo2005,Ransom2008}. Some notable clusters are Terzan 5 with 37 pulsars \citep{Cadelano2018} and 47 Tucanae which is known to host 25 pulsars, all of which have spin periods less than 8ms \citep{Freire2017}. Observations of X-ray emission from globular clusters led to the first suggestion that globular clusters could host central BHs of intermediate mass \citep[$10^3 - 10^5 M_{\odot}$][]{Silk1975,Colbert2006,Lin2018}. Moreover, extrapolation to the low-mass end of the `$M$-$\sigma$ relation' \citep{Ferrarese2000, Gebhardt2000} also suggests that globular clusters should host IMBHs \citep{Sadoun2012,Graham2019}. Suggestive evidence for IMBHs is put forth via observational radiative accretion signatures \citep[e.g.]{Ulvestad2007} and stellar kinematics \citep[e.g. ][]{Gebhardt2002, VDM2010, Feldmeier2013}, but other studies dispute this as evidence for IMBH \citep[e.g.][]{Miller2012, Baumgardt2017}. For a complete review and discussion on the current observational evidence for IMBH, see \citet{Mezcua2017}. \newline

\noindent If globular clusters and dwarf galaxies do host IMBH, then the large stellar densities and expected pulsar population make these ideal, clean environments to search for PSR-BH systems from the perspectives of testing GR. As discussed, there is expected to be a large fraction of MSP which are highly desirable from the perspective of precision astronomy, given their enhanced spin period (i.e. the clock is more precise) and their long term stability (i.e. the clock is more accurate) owing to the lack of glitches - indeed the canonical pulsars are dominated by timing noise and rotational instability, whilst MSP timing models are white noise dominated \citep{Verbiest2009} and so stand most to gain from the increased sensitivity of the next generation of radio telescopes. Moreover, whilst pulsars in globular clusters are more distant and so more faint, they are also localised in central regions, making it easier for deep radio observations, whilst the increased sensitivity of the next generation of radio telescopes will make it easier to detect their flux. 

\section{Conclusion}
\noindent In summary, we have calculated the Einstein delay and the propagation delays for a typical MSP-E/IMRB system using a fully general relativistic method which accounts for both strong-field curvature effects on the photon spacetime trajectory and the impact of extended spinning body dynamics on the pulsar orbital motion. We have then compared this with the linear, weak-field PK solution which is typically used in pulsar studies. We have shown that there exists a significant discrepancy between the post-Keplerian timing delays and the fully general relativistic solutions in strong-field regimes of gravity. Accounting for this discrepancy is essential for both detecting MSP-E/IMRB systems and using them as probes of strong-field GR.

\section{Acknowledgments}
We thank the anonymous referee for constructive comments which served to much improve the original manuscript.

%
   \bibliographystyle{aa} 
   \bibliography{refs} 

\end{document}